\documentclass[twocolumn, prl, aps, superscriptaddress,longbibliography,showpacs,amsmath,amssymb,floatfix]{revtex4-1}           
\usepackage{graphicx}                                     
\usepackage{amssymb}
\usepackage{amsmath}
\usepackage{epsfig}
\usepackage{color}
\usepackage{tabu}
\usepackage{mathtools}
\usepackage[colorlinks,linkcolor=blue,anchorcolor=blue,citecolor=blue,urlcolor=blue]{hyperref}
\usepackage{physics}
\usepackage{float}
\usepackage{diagbox}

\begin{document}

\title{Entanglement distribution in fermion model with long-range interaction}
\author{Long Xiong}
\affiliation{Key Lab of Quantum Information, Chinese Academy of Sciences, School of physics, University of Science and Technology of China, Hefei, 230026, P.R. China}
\author{Yue-Xing Huang}
\affiliation{Key Lab of Quantum Information, Chinese Academy of Sciences, School of physics, University of Science and Technology of China, Hefei, 230026, P.R. China}
\author{Yu-Chun Wu}
\affiliation{Key Lab of Quantum Information, Chinese Academy of Sciences, School of physics, University of Science and Technology of China, Hefei, 230026, P.R. China}
\affiliation{Synergetic Innovation Center of Quantum Information and Quantum Physics, University of Science and Technology of China, Hefei, 230026, P.R. China}
\author{Yong-Sheng Zhang}
\affiliation{Key Lab of Quantum Information, Chinese Academy of Sciences, School of physics, University of Science and Technology of China, Hefei, 230026, P.R. China}
\affiliation{Synergetic Innovation Center of Quantum Information and Quantum Physics, University of Science and Technology of China, Hefei, 230026, P.R. China}
\author{Guang-Can Guo}
\affiliation{Key Lab of Quantum Information, Chinese Academy of Sciences, School of physics, University of Science and Technology of China, Hefei, 230026, P.R. China}
\affiliation{Synergetic Innovation Center of Quantum Information and Quantum Physics, University of Science and Technology of China, Hefei, 230026, P.R. China}
\author{Ming Gong}
\email{gongm@ustc.edu.cn}
\affiliation{Key Lab of Quantum Information, Chinese Academy of Sciences, School of physics, University of Science and Technology of China, Hefei, 230026, P.R. China}
\affiliation{Synergetic Innovation Center of Quantum Information and Quantum Physics, University of Science and Technology of China, Hefei, 230026, P.R. China}
\date{\today }

\begin{abstract}
How two-party entanglement (TPE) is distributed in the many-body systems? This is a fundamental issue because the total TPE 
between one party with all the other parties, $\mathcal{C}^N$, is upper bounded by the Coffman, Kundu and Wootters (CKW) 
monogamy inequality, from which $\mathcal{C}^N \le \sqrt{N-1}$ can be proved by the geometric inequality. 
Here we explore the total entanglement $\mathcal{C}^\infty$ and the associated total tangle $\tau^\infty$ in a $p$-wave free 
fermion model with long-range interaction, showing that $\mathcal{C}^\infty \sim \mathcal{O}(1)$ and $\tau^\infty$ may become
vanishing small with the increasing of long-range interaction. However, we always find $\mathcal{C}^\infty \sim 2\xi \tau^\infty$, 
where $\xi$ is the truncation length of entanglement, beyond which the TPE is quickly vanished, hence $\tau^\infty \sim 1/\xi$. 
	This relation is a
direct consequence of the exponential decay of the TPE induced by the long-range interaction. These results unify 
the results in the Lipkin-Meshkov-Glick (LMG) model and Dicke model and generalize the Koashi, Buzek and Imono bound
to the quantum many-body models, 
with much broader applicability. 
\end{abstract}

\maketitle

Entanglement is essential for quantum 
information \cite{gisin2002quantum,ekert1991quantum,bennett1993teleporting,bouwmeester1997experimental} and quantum computation \cite{nielsen2002quantum,bremner2002practical,zhang2003geometric}.
In the many-body systems, entanglement can  be distributed between two arbitrary parties, which hereafter will be 
named as two-party entanglement (TPE) \cite{peres1996separability,horodecki1996separability,hill1997entanglement, wootters1998entanglement}. In general, the long-range 
interaction is required for long-distance TPE \cite{venuti2006,venuti2007,Giampaolo_2010,sahling2015experimental}.
However, the monogamy of entanglement from the Coffman, Kundu, Wootters (CKW) 
conjecture restricts the total entanglement shared between one party with all the other parties with
\cite{coffman2000distributed, hill1997entanglement, wootters1998entanglement, osborne2006general, Kim2010}
\begin{equation}
\tau^N := \sum_{i \ne 1}^N \tau(\rho_{A_1, A_i}) \le \tau( \rho_{A_1|A_2A_3\cdots A_N}),
\end{equation}
where the $N$ qubits are denoted by $A_i$, with the density matrix is denoted by $\rho_{A_1, A_2, \cdots, A_N}$. This 
inequality also holds in all multimode Gaussian states \cite{adesso2006continuous, hiroshima2007monogamy}.
Here the tangle is given by $\tau(\rho) = \mathcal{C}^2(\rho)$, where $\mathcal{C}$ is the two-party concurrence \cite{hill1997entanglement}. The right-hand side of the inequality is bounded from above by 
$\tau( \rho_{A_1|A_2A_3\cdots A_N}) \le 1$, which is the maximum entropy of a single particle density matrix measured for $A_1$.
This inequality is tight and can be satisfied by some pure states (such as the W states); see 
\cite{osborne2006general, Qu2007, Dur2000, lohmayer2006entangled, geetha2015monogamous}. By the geometric inequality
\begin{equation}
\mathcal{C}^N := \sum_{i \ne 1}^N \mathcal{C}(\rho_{A_1 A_i}) \le \sqrt{N-1} < \sqrt{N},
\label{eq-CN}
\end{equation}
in which the tightness can be achieved using some asymmetric $W$ states \cite{Qu2007, Cmax}. 
However, in the realistic many-body models, the total 
entanglement $\mathcal{C}^N$ is much smaller than the above upper bound. 
In the Lipkin-Meshkov-Glick (LMG) model with identical interaction between two arbitrary parties \cite{lipkin1965validity,
wang2002pairwise,Yin2011,dusuel2004finite,dusuel2005continuous,orus2008equivalence}, it has been found that 
$\tau(\rho_{A_1, A_i}) \sim 1/N^2$ and $\mathcal{C}(\rho_{A_1 A_i}) \sim 1/N$, thus while the total entanglement 
$\mathcal{C}^{N} \rightarrow \mathcal{O}(1) \ll \sqrt{N}$, the total tangle $\tau^N \sim 1/N$ becomes 
vanishing small as $N \rightarrow \infty$ \cite{botet1982size, dusuel2004finite, Makhalov2019}. The large $N$ limit 
is nothing but just the classical approximation of the quantum spin operator, based on which its phase transition can be solved.
These results are also found in the Dicke state, spin coherent state and spin squeezed states \cite{wang2002pairwise, Yin2011}.
For the symmetric case, the upper bound $\mathcal{C}(\rho_{A_1A_i}) \le 2/N$ is proven by Koashi, Buzek and 
Imono (KBI) \cite{koashi2000entangled}. These results raise an important yet intriguing question that how the total entanglement is
distributed in the general many-body models in the thermodynamic limit? Obviously, when the LMG and Dicke models
approach the classical limit without TPE at $N \rightarrow \infty$, it will be totally different in the other 
quantum systems with nonzero short-distant TPE. The answer to this question maybe useful for us to 
understand the role of entanglement in the many-body models and their phase transitions.

\begin{figure}[htbp!]
    \centering
    \includegraphics[width=0.46\textwidth]{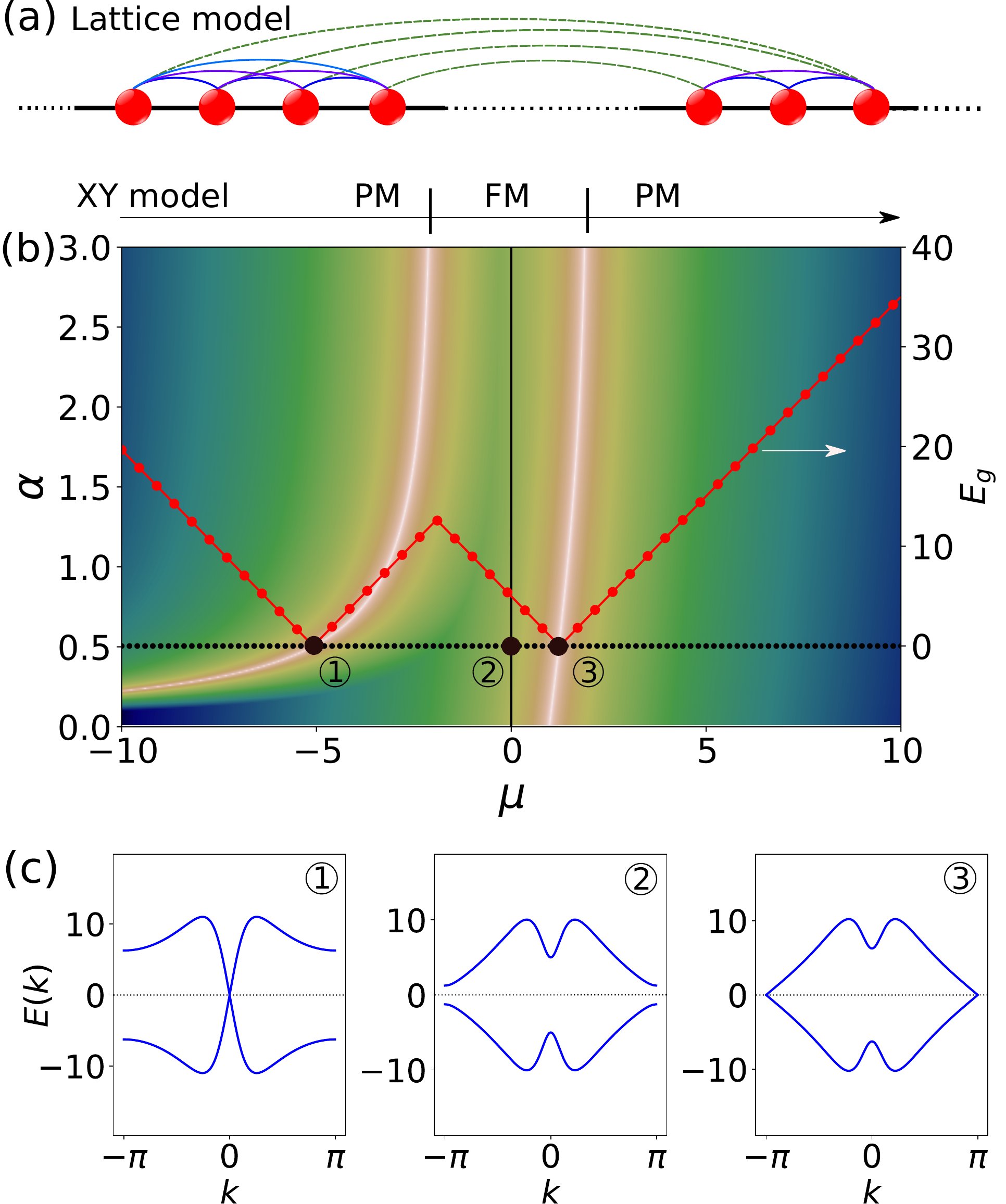}
    \caption{
	(a) One dimensional free fermion model with long-range interaction. (b) 
	The phase diagram of Eq. \ref{eq-Hamiltonian2}, with phase boundaries given by 
	Eq. \ref{eq-boundary} (see the white lines). The dotted line is the global energy gap
	$E_g = \min \{ 2E(k) \}$ at $\alpha = \beta = 0.511$, which closes and reopens at the boundaries.
	When $\alpha \rightarrow \infty$, this model is reduced to the transverse XY model with 
	phase boundaries at $\mu = \pm 2$, which corresponds to the transition from the ferromagnetic 
	(FM) phase to the paramagnetic (PM) phase. 
	(c) The spectra of the Hamiltonian at the two phase boundaries and the intermediate phase using 
	$\alpha = 0.511$, $\Delta = 1.3$ for $\mu = -5.0$, $0$, $1.25$ from left to right figures, 
	respectively.}
	\label{fig-figure1}
\end{figure}

We aim to explore the distribution of TPE restricted by the CKW monogamy inequality in the many-body 
system with long-range interaction, which may induce long-distance TPE. We will focus on an 
artificial $p$-wave superconducting model with long-range interaction in the thermodynamic limit. We find that: 
(I) Both the long-range interaction and large chemical potential are important for long-distance TPE; when the chemical 
potential is small, the TPE can not be distributed between distance qubits. 
(II) In the presence of long-distance TPE, the scaling laws of these long-distance TPE at the critical point can be 
used to diagnose the phase transition; however, the central charge from the entanglement entropy is still $c=1$ for 
free bosons; 
(III) In the condition with long-distance TPE, we always have $\mathcal{C}^\infty \sim 2 \xi \tau^\infty$ and 
$\mathcal{C}^\infty \sim \mathcal{O}(1)$, which have the same feature as the LMG model by replacing the total number of qubits $N$ using the truncation length $\xi$. These results may represents some general features of TPE in the many-body systems.

We consider the following $p$-wave superconducting model with long-range interaction (see Fig. \ref{fig-figure1} (a))
\begin{equation}                                                                                                                        H=-\mu \sum_{i=0}^\infty c_i^{\dagger}c_i+\sum_{i,j}^\infty [-t(d) c_i^{\dagger}c_{j}+\Delta(d) c_i c_{j}+ \text{h.c.}],
	\label{eq-Hamiltonian2}
\end{equation}
where $\mu$ is the chemical potential, and $t(d) = te^{-\alpha |d| + \alpha}$, $\Delta(d) = \Delta e^{-\beta |d| + \beta}$ 
are the hopping and pairing strengths between the sites $i$ and $j$ with separation $d = i-j$. 
This model is a direct generalization 
of the Kitaev chain for Majorana edge modes \cite{Kitaev2001unapired}, which has been a toy model for searching
of Majorana modes in experiments \cite{Alicea2012, Beenakker2013,Sato2016}. Ater a non-local Jordan-Wigner
transformation, it is changed to the transverse XY model, which is the limiting case of a lot of spin models 
\cite{greiter20141d,chhajed2020ising, Ercolessi2011, Xu2021}. Thus the above model may have features that can be 
found in a lot of many-body models. 

In the limit of short-range interaction $t(d) = t \delta_{d = \pm 1}$, $\Delta(d) = \Delta \delta_{d = \pm 1}$, this 
model has been studied in \cite{Kitaev2001unapired, osborne2002entanglement, osterloh2002scaling}, showing of 
only singularity in the first derivative of the nearest-neighbor TPE with some kind of 
universal scaling laws. This is a general feature of short-ranged interacting models \cite{venuti2006, venuti2007}. 
For the realization of long-distance TPE, one may need the long-range interaction, 
as evidenced by the LMG and Dicke models \cite{lipkin1965validity,wang2002pairwise,Yin2011, dusuel2004finite,
dusuel2005continuous,orus2008equivalence}. When $\alpha \rightarrow 0$ and $\beta \rightarrow 0$, it can be regarded
as a infinitely coordinated system \cite{botet1982size}, such as that in the LMG model. 
We emphasize that all the features demonstrated below do not depend on the particular forms of the long-range 
interaction. In the momentum space
\begin{equation}
	H= \sum_{k}
	\left(\begin{array}{cc}
	c_{k}^{\dagger} & c_{-k}\end{array}\right) \left(\begin{array}{cc}
	\epsilon_1  & i  \epsilon_2\\
	-i \epsilon_2  & -\epsilon_1 \\
	\end{array}\right) \left(\begin{array}{c}
	c_{k}\\
	c_{-k}^{\dagger}
	\end{array}\right),
\end{equation}
where 
\begin{eqnarray}
	\epsilon_1(k) && = \mu+t \exp(\alpha)\frac{\sinh(\alpha)-\cosh(\alpha)+\cos(k)}{\cosh(\alpha)-\cos(k)}, \\
	\epsilon_2(k) && = \Delta \exp(\beta)\frac{\sin(k)}{\cosh(\beta)-\cos(k)}.
\end{eqnarray}
The eigenvalues are
\begin{equation}
	E(k) = \pm \sqrt{\epsilon_1^2 + \epsilon_2^2}, \quad E_g = \text{min}_{k \in [-\pi, \pi]}(2E(k)).
\end{equation}
The second term defines the global energy gap used in Fig. \ref{fig-figure1} (b). 
The closing and reopening of the energy gap yields the phase boundaries at (independent of $\beta$ and $\Delta$) 
\begin{equation}
	\alpha^* =\ln(\mu / (2t \pm \mu)).
	\label{eq-boundary}
\end{equation}
For simplicity, hereafter, we let $t = 1$ as the basic energy scale of the above Hamiltonian. The phase diagram of $H$
is given in Fig. \ref{fig-figure1} (b), with typical band structure shown in Fig. \ref{fig-figure1} (c) for the two phase
transitions at $k = 0$ and $k = \pi$, respectively. At these boundaries, the spectra are linearized, indicating of 
criticality with nonzero central charge from conformal field theory.

\begin{figure}[htbp!]
    \centering
    \includegraphics[width=0.46\textwidth]{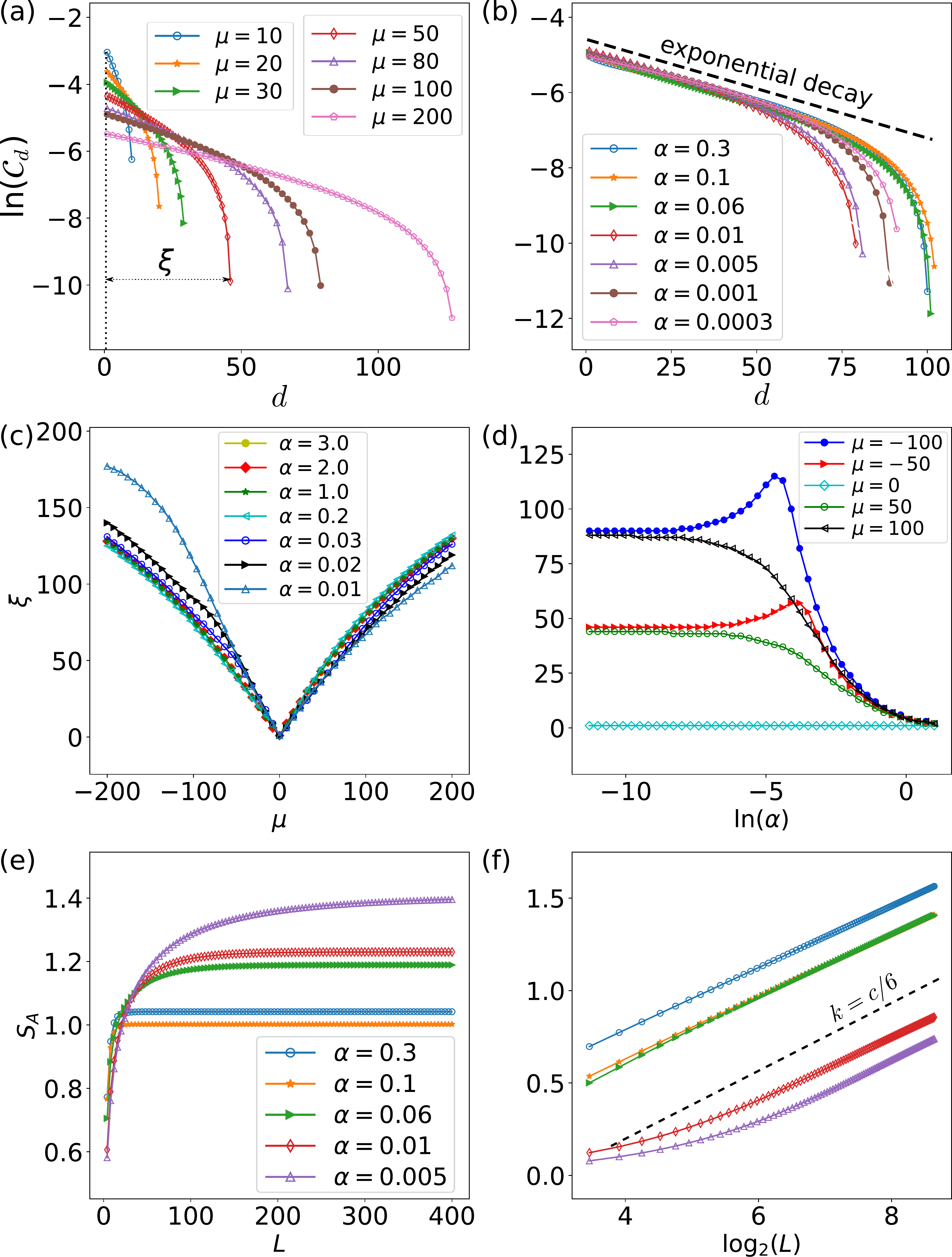}
	\caption{(a) - (b) TPE as a function of $\mu$ and $\alpha = \beta$ with fixed $\Delta = -1.0$. 
	$\xi$ is the truncation length of TPE, beyond which the TPE $\mathcal{C}_d$ will quickly drop to 
	zero; while when $|d| < \xi$, it decays exponentially with the distance $d$ (see the dashed line). 
	(c) - (d) Effect of $\mu$ and $\alpha = \beta$ on the truncation length $\xi$. 
	These two parameters are essential for the long-distance TPE. 
	(e) The entanglement entropy $S_A$ in a finite block for various $\alpha$. In the gapless regime, 
	$S_A$ scales logarithmic as a function of $\log_2 L$ as shown in (f), which is used to extract the 
	central charge $c = 1$ from the slope $k = c/6$. Saturation of $S_A$ is expected in the fully gapped phase 
	\cite{vidal2003entanglement}.}
	\label{fig-figure2}
\end{figure}

The TPE derived from the formation of entanglement \cite{wootters1998entanglement, jozsa1994fidelity} is well defined, 
termed as concurrence. We find that the two-party density matrix with separation $d$ can be obtained as 
\cite{vidal2003entanglement, pfeuty1970one, lieb1961two, jin2004quantum, latorre2003ground}
\begin{equation}
	\rho_d = \rho = \frac{1}{4} \begin{pmatrix} 
		a & 0  & 0 & f \\ 
		0	& b	& e & 0 \\ 
		0	& e^*	& b & 0 \\ 
		f^*	& 0 & 0 & d
	\end{pmatrix},
\end{equation}
where $a=P_{00}+2P_{z0}+P_{zz}$, $d=P_{00}-2P_{z0}+P_{zz}$, $b=P_{00}-P_{zz}$, $e=P_{xx}+P_{yy}$, $f=P_{xx}-P_{yy}$.
These coefficients are determined by \cite{lieb1961two,pfeuty1970one,osterloh2002scaling}
\begin{eqnarray}                                                                                                              
	&& P_{00}=1,\quad P_{zz}=\mathcal{G}_0^2-\mathcal{G}_{d}\mathcal{G}_{-d},\quad P_{z0}=P_{0z}=-\mathcal{G}_{0}, \nonumber \\
	&& P_{xx}=\text{det} (T(\mathcal{G}_{i-j-1})), \quad P_{yy}=\text{det} (T(\mathcal{G}_{i-j+1})), 
\end{eqnarray}
where $T$ is a $d\times d$ Toplitz matrix with entries $T_{i,j} = \mathcal{G}_{i-j-k}$, with 
\begin{equation}
	\mathcal{G}_{x} = \int_{-\pi}^{\pi} {\epsilon_2 \cos(kx) - \epsilon_1 \sin(kx) \over 2\pi\sqrt{\epsilon_1^2+\epsilon_2^2}}dk.
\end{equation}
The concurrence of this density matrix reads as \cite{hill1997entanglement, wootters1998entanglement}
\begin{equation}
	\mathcal{C}_d(\rho_d) = \text{max}\{0, \frac{1}{2}(|f| - |b|), \frac{1}{2}(|e| - \sqrt{ad})\}.
	\label{eq-concurrence}
\end{equation}
Due to the translational symmetry in $H$, it is sufficient to characterize the total entanglement 
between one arbitrary party with all the other parties using
\begin{eqnarray}
	\mathcal{C}^N = 2\sum_{d=1}^N \mathcal{C}_d(\rho_d), \quad \tau^N = 2\sum_{d=1}^N \tau_d(\rho_d),
	\label{eq-sumC}
\end{eqnarray}
where $\tau_d=\mathcal{C}_d^2$ and the factor 2 comes from the inversion symmetry of 
$\mathcal{C}_{-d} = \mathcal{C}_d$. 
We find a truncation length $\xi$, which means the farthest distance $d$
that the TPE can be achieved. This distance is well-defined for the reason that when the
distance $|d| > \xi$, $\mathcal{C}_d$ will quickly approach zero (see Fig. 
\ref{fig-figure2} (a) - (b)). When $d \ll \xi$, $\mathcal{C}_d$ will decay exponentially with the increase
of separation. This is in stark contrast to the model with short-range interaction \cite{osterloh2002scaling, osborne2002entanglement}, in which $\xi \sim 1/3$,
thus only the nearest-neighbor sites have significant TPE. Moreover, only the entanglement of 
$\mathcal{C}_{\pm 1}$ exhibits singularity at the critical boundaries, which has been used 
to characterize the phase transitions \cite{osterloh2002scaling,amico2006entanglement}. 
Hence we can define the total entanglement shared between one party with all the other parties 
as $\mathcal{C}^{\infty} = 2\sum_{d=1}^{\infty} \mathcal{C}_d$ and $\tau^{\infty} = 2\sum_{d=1}^{\infty} \tau_d$.

The long-range interaction is necessary but not sufficient to induce long-distance TPE. 
The truncation length $\xi$ as a function of $\mu$ and $\alpha$ (assuming $\alpha = \beta$) 
are shown in Fig. 
\ref{fig-figure2} (c) - (d), showing that when $\mu = 0$, $\xi$ will approach a small value 
$\xi \sim 1$. On the other hand, for fixed $\mu$, $\xi$ will also approach 
a constant independent of $\mu$ when $\alpha \rightarrow 0$ (the long-range limit), for the reason of divergence of 
$\epsilon_1$, $\epsilon_2$ and the energy gap $E_g$ at $\alpha \rightarrow 0$. However, we emphasize that while the 
long-range interaction can induce long-distance TPE, at the phase boundary the many-body entanglement measured using 
the Shannon entropy is not dramatically affected. By fitting the entanglement entropy using 
$S_A = \frac{c}{6} \log_2 L + s_0$, we obtain $c = 1$ \cite{vidal2003entanglement,latorre2003ground,jin2004quantum}; 
however, the longer the interaction is, the longer the length $L$ is required to obtain this perfect linear relation 
(see Fig. \ref{fig-figure2} (f)).

\begin{figure}[htbp!]
    \centering
    \includegraphics[width=0.46\textwidth]{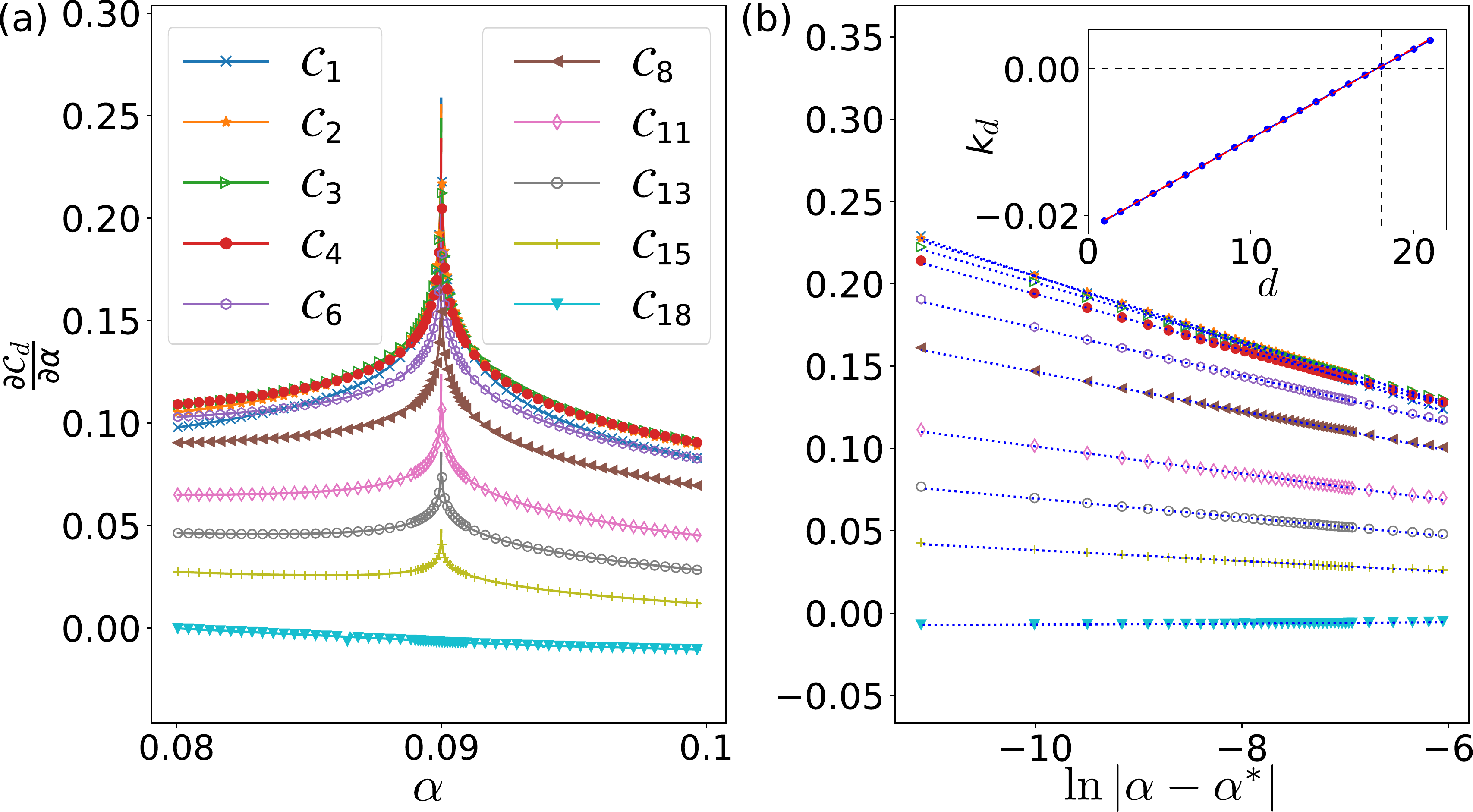}
    \caption{Universal scaling  of TPE induced by the long-range interaction $\alpha$. 
	(a) The first derivation of $\mathcal{C}_d$ as a function of $\alpha$, with critical point 
	at $\alpha^* = 0.09$  (using Eq. \ref{eq-boundary}). Here not only the nearest-neighbor TPE, but also the 
	long-distance TPE, diverge at the same $\alpha^*$. (b) $\partial \mathcal{C}_d/\partial \alpha$
	as a function of $\ln|\alpha - \alpha^*|$ for various separation $d$. Inset shows $k_d$ as a function
	of $d$, with $q = 0.0012$ and $q' = -0.022$. In all figures $\mu = -23.23722$, $\Delta = 1.3$ and 
	$\beta = 0.027$. }
    \label{fig-figure3}
\end{figure}

The long-range interaction can induce long-distance TPE, which exhibit some universal scaling laws in 
their first derivative. We want to show that (see Fig. \ref{fig-figure3}) \cite{osborne2002entanglement,gu2005ground}
\begin{eqnarray}
	\frac{\partial \mathcal{C}_d }{d \alpha} \simeq k_d \ln(|\alpha-\alpha^{*}|), \quad k_d = q |d| + q',
	\label{eq-kd}
\end{eqnarray}
where $\alpha^{*}$ is the critical values. This relation is also true as a function of the other parameters. 
The interesting point is that the long-range interaction can 
induce not only short-range TPE (by $\mathcal{C}_1$), but also long-distance TPE (by $\mathcal{C}_d$ for $|d| > 1$),
which is strictly forbidden in the transverse XY model \cite{osborne2002entanglement}. The slop $k_d$ depends 
strongly on the value of $d$ in a linear manner (inset of Fig. \ref{fig-figure3} (b)). 
The same scaling form can be obtained as a function of chemical potential with critical $\mu^*$. 
A detailed analysis shows that the above logarithmic divergence arises from the logarithmic divergence of 
$\partial \mathcal{G}_d /\partial \alpha$, which is independent of the particular forms of long-range interaction.

\begin{figure}[t!]
    \centering
    \includegraphics[width=0.46\textwidth]{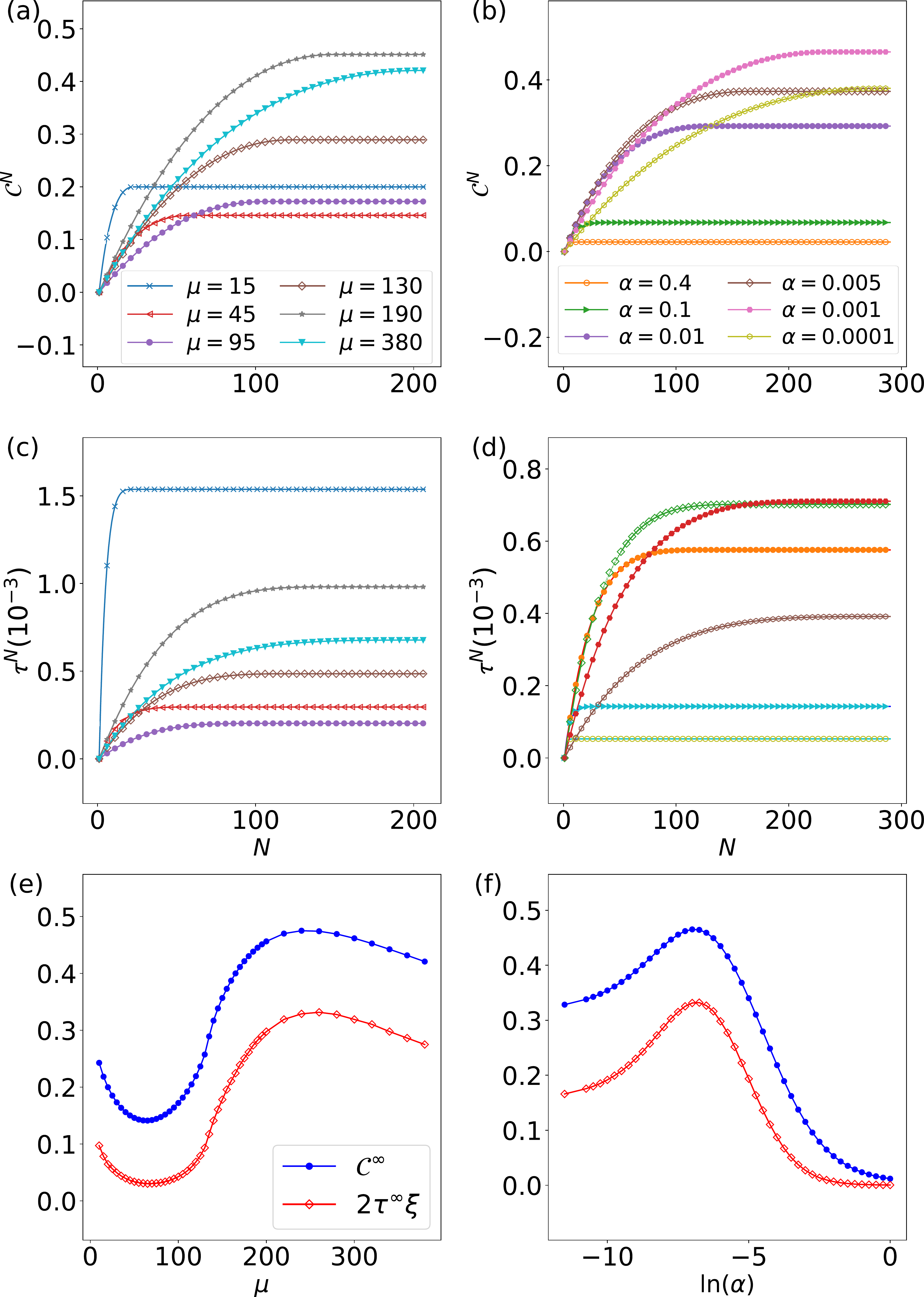}
	\caption{(a) - (b) The total entanglement $\mathcal{C}^N$ as a function of $N$, with fixed $\alpha = \beta = 0.015$. 
	(c) - (d) The corresponding total tangle $\tau^N$ for the above two cases. 
	(e) - (f) Relation between the total entanglement $\mathcal{C}^\infty$, total tangle $\tau^\infty$ and the 
	truncation length of entanglement $\xi$, showing of Eq. \ref{eq-cxi}. In (e) $\alpha = \beta = 0.015$; and
	(f) $\mu = 100$. In all figures $\Delta = 1.3$. }
    \label{fig-figure4}
\end{figure}

The upper bounds of CKW and KBI inequalities yield some dilemma to the total entanglement in the thermodynamic limit. 
These relations in the few-party systems (using the entangled photon pairs) have been 
intensively verified \cite{Zhu2020,Farias2012, Aguilar2014, Friis2019, Genovese2005}; yet in the many-body quantum 
systems they are rarely discussed. In \onlinecite{dusuel2004finite}, the concurrence in the 
LMG model has been studied using $H=-\frac{\lambda}{N} \sum_{i<j}(\sigma_x^i \sigma_x^j+\gamma \sigma_y^i \sigma_y^j) - \sum_i \sigma_z^i$, finding ($\lambda < 1$)
\begin{equation}
	\mathcal{C}_d=\frac{1-\sqrt{\frac{1-\lambda} {1-\gamma \lambda}}}{N-1}, \quad \tau^N \sim {1 \over N}. 
	\quad \mathcal{C}^N \sim N \tau^N.
	\label{eq-lmg}
\end{equation}
This model can be derived from the Dicke model \cite{Morrison2008, emary_chaos_2003, emary_quantum_2003}. 
In \cite{wang2002pairwise,Yin2011}, Wang {\it et. al.} have found the same scaling relation that $\mathcal{C}_d \sim 1/N$ 
in the Dicke state, spin coherent state and spin squeezed state (see [\onlinecite{Remarks}]). This is a somewhat expected
feature since the phase transition in the these models can be understood using classical approximation \cite{botet1982size, 
dusuel2004finite, Makhalov2019}, hence
the entanglement should be vanishing small. Then what will happen in the quantum models? We will provide a possible 
solution to the above dilemma for the quantum models, which meanwhile satisfies the CKW conjecture in the thermodynamic limit.

In Fig. \ref{fig-figure4} (a) - (d), we calculate the total entanglement and total tangle as a function of $N$. In all the
cases, due to the truncation of entanglement $\xi$, these two qualities are saturated to a finite value. However, while 
the total concurrence is significant, the total tangle is very small. 
One of the central results, stimulated by Eq. \ref{eq-lmg}, is that 
\begin{equation}
	C^\infty \sim 2\tau^\infty \xi, \quad \tau^\infty \sim \mathcal{C}_1 \sim 1/\xi. 
	\label{eq-cxi}
\end{equation}
This relation can be regarded as a generalized KBI bound \cite{koashi2000entangled}, replacying $N$ by $\xi$. 
It is a direct consequence of exponential decaying of TPE with respect to $d$, assuming $\mathcal{C}_d = 
\mathcal{C}_1  e^{-|d|/\xi}$ (see Fig. \ref{fig-figure2} (a) - (b)). 
Using the upper bound of $\tau^\infty$ from the CKW inequality, we have $\mathcal{C}_1 \le 1/\sqrt{\xi}$. 
The actual condition is, we find $\mathcal{C}^\infty \sim \mathcal{O}(1)$, thus $\mathcal{C}_1 < 1/\xi$, and 
$\tau^\infty \sim 1/\xi$. Hence the bigger $\xi$ is, the smaller the tangle $\tau^\infty$ to be. This relation unifies 
Eq. \ref{eq-lmg} in the LMG model and Dicke model by replacing $N$ with $\xi$ and generalizes its validity to 
much more quantum many-body models.

To conclude, we examine the distribution of TPE in a fermion model with long-range interaction, which is upper bounded 
by the CKW monogamy inequality. We show that the long-range interaction induces long distant TPE, all of which can 
exhibit universal scaling laws at the critical boundaries. A finite truncation length $\xi$ is required to describe the 
TPE. Based on this quantity, we find a general relation $\mathcal{C}^\infty \sim 2 \xi \tau^\infty \sim 
\mathcal{O}(1)$, which holds in both the gapped and gapless phases. Our new relation is a 
generalized KBI bound, which has much braoder applicability in the quantum many-body models.

{\it Acknowledgments}: This work is supported by the National Key Research and Development Program in China (
Grants No. 2017YFA0304504 and No. 2017YFA0304103), the strategic Priority Research Program(B) of the Chinese 
Academy of Sciences (Grant No. XDB01030100 and No. XDB01030300) and the National Natural Science Foundation of China 
(NSFC) with No. 11774328, No. 11674306 and No. 92065113. 


\begin{thebibliography}{61}
\bibitem [{Cma()}]{Cmax}%
  \BibitemOpen
  \href@noop {} {}\bibinfo {note} {For the W state $|W\rangle =
  \alpha_1|1000\cdots\rangle + \alpha_2 |0100\cdots\rangle + \cdots +
  \alpha_N|\cdots 0001\rangle$, where $\alpha_i \in \mathbb{R}$ and $\sum_i
  |\alpha_i|^2 = 1$ \cite{Qu2007}, we have $\mathcal{C}_{A_1 A_i}^2 =
  4\alpha_i^2 \alpha_i^2$, and $\mathcal{C}_{A_1|A_2A_3\cdots A_N}^2 =
  4\alpha_1^2 (\sum_{i=2}^N \alpha_i^2)$. We can prove that the upper bound of
  Eq. 2 with $\mathcal{C}^N =\sqrt{N-1}$ in the main text can be achieved by
  $\alpha_1 = 1/\sqrt{2}$ and $\alpha_i = 1/\sqrt{2(N-1)}$ for $i = 2$,
  $\cdots$, $N$, which is an asymmetric state. When all $\alpha_i = 1/\sqrt{N}$
  for a pure symmetric state, the total entropy will be immediately reduced to
  $\mathcal{C}^{N} = 2(N-1)/N < 2$. This upper bound is rigorously proven by
  Koashi, Buzek and Imono \cite{koashi2000entangled}, assuming that in a
  symmetric $N$ qubits system, the two-party density matrix $\rho_{A_iA_j}$
  ($i\ne j$) is independent of the choice of the pair $i$ and $j$.}\BibitemShut
  {Stop}%
\end{thebibliography}

%

\end{document}